\documentclass{article}

\usepackage{arxiv}

\usepackage{amsmath}
\usepackage[utf8]{inputenc} 
\usepackage[T1]{fontenc}    
\usepackage{hyperref}       
\usepackage{url}            
\usepackage{booktabs}       
\usepackage{amsfonts}       
\usepackage{nicefrac}       
\usepackage{microtype}      
\usepackage{cleveref}       
\usepackage{lipsum}         
\usepackage{graphicx}
\usepackage{natbib}
\usepackage{doi}

\DeclareUnicodeCharacter{2061}{}

\title{A (1.999999)-approximation ratio for\\vertex cover problem}

\author{ \href{https://orcid.org/0000-0002-5211-2413}{\includegraphics[scale=0.06]{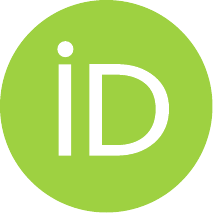}\hspace{1mm}Majid ~Zohrehbandian}
\\
	Department of Mathematics, Karaj Branch, 
    \\Islamic Azad University, Karaj, Iran. \\
	\texttt{zohrebandian@yahoo.com} \\
}

\renewcommand{\headeright}{M. Zohrehbandian}
\renewcommand{\shorttitle}{A (1.999999)-approximation ratio for vertex cover problem}

\hypersetup{
pdftitle={Less than 2 approximation for VCP},
pdfsubject={cs.CC},
pdfauthor={Majid ~Zohrehbandian},
pdfkeywords={First keyword, Second keyword, More},
}

\begin{document}
\maketitle

\begin{abstract}
The vertex cover problem is a famous combinatorial problem, and its complexity has been heavily studied. While a 2-approximation can be trivially obtained for it, researchers have not been able to approximate it better than 2-\textit{o}(1). 
In this paper, by introducing a new semidefinite programming formulation that satisfies new properties, we introduce an approximation algorithm for the vertex cover problem with a performance ratio of 1.999999 on arbitrary graphs, en route to answering an open question about the correctness of the unique games conjecture.\end{abstract}



\section{Introduction}

In complexity theory, the abbreviation \textit{NP} refers to "nondeterministic polynomial", where a problem is in \textit{NP} if we can quickly 
(in polynomial time) test whether a solution is correct. \textit{P} and \textit{NP}-complete problems are subsets of \textit{NP} problems. 
We can solve \textit{P} problems in polynomial time while determining whether or not it is possible to solve \textit{NP}-complete problems quickly 
(called the \textit{P vs NP} problem) is one of the principal unsolved problems in Mathematics and Computer science.

Here, we consider the vertex cover problem (VCP), a famous \textit{NP}-complete problem, which cannot be approximated within a factor of 1.36 [1], 
unless \textit{P}\ =\ \textit{NP}. In contrast, a 2-approximation factor can be trivially obtained by taking all the vertices of a maximal matching 
in the graph. However, improving this simple 2-approximation algorithm is hard [2, 3]. 

In this paper, based on a lower bound on the objective value of VCP feasible solutions, 
we introduce a (2-$\varepsilon$)-approximation ratio, where the value of $\varepsilon$ is not constant.
Then, we introduce a new semidefinite programming (SDP) formulation and fix the $\varepsilon$ value equal to $\varepsilon$=0.000001, 
to produce a 1.999999-approximation ratio on arbitrary graphs. 

The rest of the paper is structured as follows. 
Section 2 is about the vertex cover problem and introduces new properties. 
In section 3, using a new SDP model whose solution satisfies the properties, we propose a solution algorithm for VCP 
with a performance ratio of 1.999999 on arbitrary graphs. 
Finally, Section 4 concludes the paper.

\section{Performance ratio based on VCP feasible solutions}

In the mathematical discipline of graph theory, a vertex cover of a graph is a set of vertices such that 
each edge of the graph is incident to at least one vertex of the set. The problem of finding a minimum vertex cover 
is a typical example of an \textit{NP}-complete optimization problem. In this section, 
we calculate the performance ratios of VCP feasible solutions to produce an approximation ratio of 2-$\varepsilon$, 
where the value of $\varepsilon$ is not constant and depends on the VCP objective value. 
Then, in the next section, we will fix the value of $\varepsilon$ equal to $\varepsilon$=0.000001, to produce a 1.999999-approximation ratio 
for the vertex cover problem.

Let $G=(V,E)$ be an undirected graph on vertex set $V$ and edge set $E$, where $\mid$V$\mid =n$. 
Throughout this paper, $z^*(G)$ is the optimal value for the vertex cover problem on $G$, and 
VCP feasible solutions have been introduced by a vertex partitioning $V=V_1\cup V_0$ with an objective value $\mid V_1\mid$. The integer linear programming (ILP) model for VCP is as follows; i.e. $z1^*=z^*(G)$.\\
$$(1)\ min_{s.t.}\ ⁡z1=\sum_{i\in V}x_{i}$$
$$x_{i}+x_{j}\geq 1\ \ \ ij\in E$$
$$x_{i}\in \{0,+1\}\ \ \ i\in V$$
\textbf{Lemma 1. [4]} Let $x^*$ be an extreme optimal solution to the linear programming (LP) relaxation of the model (1). 
Then $x_j^*\in \{0,0.5,1\}$ for $j\in V$. If we define $V^0=\{j\in V\mid x_j^*=0\}$, $V^{0.5}=\{j\in V\mid x_j^*=0.5\}$ and 
$V^1=\{j\in V\mid x_j^*=1\}$, then there exists a VCP optimal solution which includes all of the vertices $V^1$, and it is a subset of $V^{0.5}\cup V^1$.
\\
\\
\textbf{Theorem 1. } Let $x^*$ be an extreme optimal solution to the LP relaxation of the model (1), 
$V^0=\{j\in V\mid x_j^*=0\}$, $V^{0.5}=\{j\in V\mid x_j^*=0.5\}$, $V^1=\{j\in V\mid x_j^*=1\}$, and $G_{0.5}$ 
be the induced graph on the vertices $V^{0.5}$. If we can introduce a vertex cover feasible partitioning 
$V^{0.5}=V_1^{0.5}\cup V_0^{0.5}$ with an approximation ratio of $1\leq \rho <2$, for the VCP on $G_{0.5}$, 
then the vertex cover feasible partitioning $V=(V_1\cup V_0)=(V_1^{0.5}\cup V^1)\cup (V_0^{0.5}\cup V^0)$, 
has an approximation ratio of $1\leq \rho <2$, for the VCP on $G$.
\\
\textbf{Proof. } Based on the approximation ratio of $\frac{\mid V_1^{0.5}\mid}{z^*(G_{0.5})}\leq\rho$, we have, 
$$\mid V_1^{0.5}\mid + \mid V^1\mid \leq \rho z^*(G_{0.5}) + \rho \mid V^1\mid$$
Therefore,  
$\frac{\mid V_1\mid}{z^*(G)}=\frac{\mid V_1^{0.5}\mid+\mid V^1\mid }{z^*(G_{0.5})+\mid V^1\mid }\leq\rho\ \diamond$
\\

Based on the Theorem (1), it is sufficient to produce an approximation ratio of $1\leq \rho <2$, on $G_{0.5}$. 
Then, let's focus on $G_{0.5}$ and assume that for the optimal solution of the LP relaxation of the model (1), we have 
$V^0=V^1=\{\}$, $V^{0.5}=V$; i.e. $G=G_{0.5}$. 
\\

We know that we can efficiently solve the following SDP formulation, 
as a relaxation of the VCP model (1).
$$(2)\ min_{s.t.}\ ⁡z2=\sum_{i\in V} X_{oi}$$
$$X_{oi}+X_{oj}-X_{ij}=1\ \ \ ij\in E$$
$$X_{ii}=1,\ \ \ 0\leq X_{ij}\leq +1\ \ \ i,j\in V\cup\{o\}$$
$$X\succeq 0$$
Moreover, by introducing the normal vectors $v_o,v_1,...,v_n$, the SDP model (2) can be written as follows, 
where $v_iv_j=X_{ij}$, and $V_1=\{i\in V\mid v_i = v_o\}$ is a feasible vertex cover, 
and $V_o=V-V_1$ is the set of perpendicular vectors to $v_o$. 
\\
\\
$$(3)\ min_{s.t.}\ ⁡z3=\sum_{i\in V} v_ov_i$$
$$v_ov_i+v_ov_j-v_iv_j=1\ \ \ ij\in E$$
$$v_iv_i=1,\ \ \ 0\leq v_iv_j\leq +1\ \ \ i,j\in V\cup\{o\}$$
\\
\textbf{Theorem 2.} Let $\frac{n}{2}+\frac{n}{k}$ be a lower bound on VCP optimal value; i.e. $z^*(G)\geq\frac{n}{2}+\frac{n}{k}=\frac{(k+2)n}{2k}$. Then, all vertex cover feasible partitioning $V=V_1\cup V_0$ satisfy the approximation ratio  
$\frac{\mid V_1 \mid }{z^*(G)}\leq \frac{2k}{k+2}<2$.
\\
\textbf{Proof.} If $z^*(G)\geq \frac{(k+2)n}{2k}$, then $\frac{n}{z^*(G)}\leq \frac{2k}{k+2}$. 
Therefore, 
$$\frac{\mid V_1 \mid}{z^*(G)}\leq\frac{n}{z^*(G)}\leq\frac{2k}{k+2}<2$$ 
and this completes the Proof $\diamond$
\\
\\
\textbf{Theorem 3.} Let $z^*(G)\geq\frac{n}{2}$, and $V=V_1\cup V_0$ is a VCP feasible partitioning, 
where $\mid V_1 \mid \leq \frac{kn}{k+1}$ and $\mid V_0 \mid \geq\frac{n}{k+1}$ (or $\mid V_1 \mid\leq k\mid V_0 \mid$). Based on such a solution, we have an approximation ratio $\frac{\mid V_1 \mid}{z^*(G)}\leq\frac{2k}{k+1}<2$.
\\
\textbf{Proof.} If $\mid V_1 \mid \leq\frac{kn}{k+1}$, then $n\geq\frac{k+1}{k} \mid V_1 \mid$. 
Hence, $z^*(G)\geq\frac{n}{2}\geq\frac{k+1}{2k} \mid V_1 \mid$ and $\frac{\mid V_1 \mid}{z^*(G)}\leq\frac{2k}{k+1}<2\ \diamond$
\\

In the next section, we will introduce a new SDP model to find a suitable lower bound or feasible solution and apply Theorem (2) or Theorem (3). 

\section{A (1.999999)-approximation algorithm on arbitrary graphs}

In section 2, we introduced a (2-$\varepsilon$)-approximation ratio for VCP,
where $\varepsilon$ value was not constant. In this section, we fix the value of $\varepsilon$ equal to $\varepsilon$=0.000001 
to produce a 1.999999-approximation ratio on arbitrary graphs. To do this, 
we introduce the following property on the solution of the SDP model (3).
\\
\\
\textbf{Property 1.} For some vertex cover problems, the optimal solution of the  SDP model (3) satisfies the following conditions. 
\\
\textbf{a)} $\mid\{j\in V:\ v_o^*v_j^*\ <\ 0.5\}\mid < 0.000001n$. 
\\
\textbf{b)} $\mid\{j\in V:\ v_o^*v_j^*\ >\ 0.5004\}\mid < 0.01n$. 
\\
\\
\textbf{Theorem 4.} If $z^*(G)\geq\frac{n}{2}$ and the optimal solution of the SDP model (3) does not meet the Property (1), 
then we can produce a VCP solution with a performance ratio of 1.999999.
\\
\textbf{Proof.} If the optimal solution of the SDP model (3) does not meet the Property (1.a), 
then we can introduce $V_0=\{j\in V \mid v_o^*v_j^*<0.5\}$ and $V_1=V-V_0$, to have a VCP feasible solution 
with $\mid V_0 \mid\geq 0.000001n$ and $\mid V_1 \mid\leq 0.999999n\leq 999999\mid V_0 \mid$. 
Therefore, for such a solution and based on the Theorem (3), we have an approximation ratio of 
$\frac{\mid V_1 \mid}{z^*(G)}<\frac{2(999999)}{999999+1}=1.999998<1.999999$.

Otherwise, if the optimal solution of the SDP model (3) meets the Property (1.a) but does not meet the Property (1.b), then there exists the following lower bound on $z^*(G)$ value.   
$$z^*(G)\geq z3^*\geq (0)(0.000001n)_{\{s.t.\ v_o^*v_j^*<0.5\}}$$
$$+(0.5)(0.989999n)_{\{s.t.\ v_o^*v_j^*\geq 0.5\}}+(0.5004)(0.01n)_{\{s.t.\ v_o^*v_j^*>0.5004\}}$$ 
$$=\frac{n}{2}+0.0000035n$$

Note that, the Property (1.a) is met and we have less than 0.000001n of vertices $j\in V$ with $v_o^*v_j^*<0.5$.  
The Property (1.b) is not met and we have more than 0.01n of vertices $j\in V$ with $v_o^*v_j^*>0.5004$. 
Therefore, in the above inequality, the first summation is the lower bound on the vertices $j\in V$ with $v_o^*v_j^*<0.5$, 
and the third summation is the lower bound on only 0.01n of the vertices $j\in V$ with $v_o^*v_j^*>0.5004$. In other words, 
beyond the 0.01n of such vertices are considered in the second summation. 
Moreover, the second summation is the lower bound on the other vertices (the vertices $j\in V$ with $0.5\leq v_o^*v_j^*\leq 0.5004$, and the vertices $j\in V$ with $v_o^*v_j^*>0.5004$ beyond the 0.01n of such vertices considered in the third summation).

Therefore, based on the above lower bound on $z^*(G)$ value and based on the Theorem (2), all VCP feasible solutions $V=V_1\cup V_0$ satisfy the approximation ratio $\frac{\mid V_1 \mid}{z^*(G)}\leq\frac{2(\frac{1}{0.0000035})}{\frac{1}{0.0000035}+2}<1.999999\ \diamond$
\\
\\
\textbf{Definition 1.} Let $\varepsilon$=0.0004, and $V_\varepsilon=\{j\in V \mid 0.5\leq v_o^*v_j^*\leq 0.5+\varepsilon\}$, and $E_\varepsilon =\{ij\in E \mid i,j\in V_\varepsilon \}$.
\\

After solving the SDP model (3) on problems with $z^*(G)\geq\frac{n}{2}$, 
\\
- If the SDP model (3) solution does not meet the Property (1), we have a performance ratio of 1.999999.\\ 
- Otherwise, $\mid V_\varepsilon \mid\geq 0.989999n$, and $0\leq v_i^*v_j^*\leq 2\varepsilon\ \ (ij\in E_\varepsilon$); i.e. the corresponding vectors of the edges in $E_\varepsilon$ are almost perpendicular to each other.
\\

Therefore, to produce a VCP performance ratio of 1.999999 for problems with $z^*(G)\geq\frac{n}{2}$, we need a solution for the SDP model (3) that does not meet the Property (1). To do this, we will introduce a new SDP model, whose optimal solution does not meet the Property (1) unless the induced graph on $V_\varepsilon$ is bipartite. In other words, we want to approximately satisfy the conditions of the following theorem while still being able to use the result of the theorem.
\\
\\
\textbf{Theorem 5.}  It is impossible to have $2t+1$ normalized vectors $v_1,...,v_{2t+1}$, where  $v_jv_{j+1}=0$, and $v_j+v_{j+1}=v_{2t+1}+v_1$ ($j=1,...,2t$), and $v_{2t+1}v_1=0$. 
\\
\textbf{Proof.} Let $U=v_{2t+1}+v_1$. Then $\mid U \mid=\sqrt{2}$, and $Uv_j=1$ ($j=1,...,2t+1$). 
$$(v_1+v_2)+(v_2+v_3)+...+(v_{2t}+v_{2t+1})+(v_{2t+1}+v_1)=(2t+1)U$$
Hence, $$2(v_1+...+v_{2t}+v_{2t+1})=(2t+1)U$$
and $$(v_1+v_2)+(v_3+v_4)+...+(v_{2t-1}+v_{2t})+v_{2t+1}=(t+0.5)U$$
Therefore, $tU+v_{2t+1}=(t+0.5)U$, and this concludes that $v_{2t+1}=0.5U$ which is a contradiction $\diamond$
\\
\\
\textbf{Clime 1.} It is impossible to have $2t+1$ normalized vectors $v_1,...,v_{2t+1}$, where all consecutive vectors $v_j$, $v_{j+1}$ ($j=1,...,2t$), and $v_{2t+1}$, $v_1$ are almost perpendicular to each other, and $v_j+v_{j+1}$ ($j=1,...,2t$) and $v_{2t+1}+v_1$ are almost equal to a vector $U$ with $\mid U \mid=\sqrt{2}$, and $Uv_j=1$ ($j=1,...,2t+1$). 
\\
\textbf{Proof.} Read the rest.  
\\

Let $G2=(V_{new},E_{new})$ be a new graph, where we add two adjacent vertices $a$ and $b$ to the graph $G$, 
and connect all vertices of $G$ to them. 
Then, based on the SDP model (3), and by introducing the normal vectors $v_o,v_1,...,v_n,v_a,v_b$, we introduce a new SDP model as follows, where $V_{1new}=V_1=\{i\in V_{new}\mid v_i = v_o\}$ corresponds to a feasible vertex cover on graph $G$, and  
$V_{0new}=V_0=V-V_1$ corresponds to perpendicular vectors to $v_o$. 
\\
\\
\\
\\
$$(4)\ min_{s.t.}\ ⁡z4=\sum_{i\in V} v_ov_i$$
$$SDP\ (3)\ constraints\ on\ G$$
$$v_ov_i+v_ov_j-v_iv_j=1\ \ \ i\in V,\ j\in \{a,b\}$$
$$-0.5\leq v_iv_j\leq +0.5\ \ \ i\in V,\ j\in \{a,b\}$$
$$v_iv_i=1,\ \ \ v_ov_i=\ +0.5\ \ \ i\in \{a,b\}$$
$$v_av_b=\ 0$$
\\
\textbf{Lemma 2.} Due to the additional constraints, we have $z4^*\geq z3^*$. Moreover, to produce a feasible solution 
for the SDP model (4) on $G2$, we can add suitable vectors $v_a$ and $v_b$ to each VCP feasible partitioning $V=V_1\cup V_0$ on $G$, where 
$v_iv_j=\ +0.5$ for $i\in V_1,\ j\in \{a,b\}$, and
$v_iv_j=\ -0.5$ for $i\in V_0,\ j\in \{a,b\}$ 
(For example, for $v_o=v_i=[0.5,0.5,0.5,0.5]^t\in V_1$ and $v_i=[-0.5,-0.5,0.5,0.5]^t\in V_0$, we can introduce $v_a=e_1=[1,0,0,0]^t$, and $v_b=e_2=[0,1,0,0]^t$). Therefore, $z4^*\leq z^*(G)$.
\\

We are going to prove that by solving the SDP model (4) on problems with $z^*(G)\geq\frac{n}{2}$, 
it is impossible to produce a solution that meets the Property (1) on $G$, unless the induced graph on $V_\varepsilon$ is bipartite.
\\
\\
\textbf{Theorem 6.} For four normalized vectors $v_1,v_2,v_3,v_4$ which are perpendicular to each other, 
there exists exactly one normalized vector $v$ with $vv_i=0.5\ (i=1,2,3,4)$. Such a vector $v$ satisfies the equation $v=0.5(v_1+v_2+v_3+v_4)$.
\\
\textbf{Proof.}

Due to $v_1v_2=0$, we have $\mid v_1+v_2 \mid=\sqrt{\mid v_1 \mid^2+\mid v_2 \mid^2}=\sqrt{2}$.

Due to $v_3v_4=0$, we have $\mid v_3+v_4 \mid=\sqrt{\mid v_3 \mid^2+\mid v_4 \mid^2}=\sqrt{2}$.

Due to $(v_1+v_2)(v_3+v_4)=0$, we have $\mid v_1+v_2+v_3+v_4 \mid=2$.

Moreover, we have $(v_1+v_2+v_3+v_4)v=2$. Hence, $\mid v_1+v_2+v_3+v_4 \mid\mid v \mid cos⁡(\theta)=2$ and 
this concludes that $\theta=0$ and $v=0.5(v_1+v_2+v_3+v_4)\ \diamond$
\\
\\
\textbf{proposition 1.} Let $w=0.5(v_1+v_2+v_3+v_4)$, for four normalized vectors $v_1,v_2,v_3,v_4$ which are almost perpendicular to each other. Then, a normalized vector $v$ with $0.5\leq vv_i\leq 0.5+\varepsilon\ (i=1,2,3,4)$ is almost equal to $w$. In other words, there exists a vector $r$ with $w+r=v$, where $-\varepsilon\le \mid w\mid-\mid v\mid\le\varepsilon,\ \mid r\mid\leq \varepsilon$, and $cos(v,w)\geq 1-\varepsilon$.
\\
\\
\textbf{Theorem 7.} Let $\theta(v,w)=cos^{-1}(
\frac{v.w}{\mid v\mid\mid w\mid}
)$. For $n+1$ vectors $v_o,v_1,...,v_n$ with $0^o\le\theta(v_i,v_j)\le 90^o$ ($i,j=o,1,...,n$), we have $$\theta(v_o,\sum_{i=1}^nv_i)\le max\{\theta(v_o,v_i):\ i=1,...,n\}$$
In other words, the angle between the  vectors $v_o$ and $w=\sum_{i=1}^nv_i$ is smaller than the maximum angle between the pair of vectors $v_o$ and $v_i\ (i=1,...,n)$.
\\
\textbf{Proof.} 
We give proof by induction on n. For $n=3$
vectors $v_o,v_1,v_2$, if $\theta(v_o,v_1+v_2)>max\{\theta(v_o,v_1),\theta(v_o,v_2)\}$ then $\theta(v_o,v_1+v_2)>\theta(v_o,v_1)$ and 
$\theta(v_o,v_1+v_2)>\theta(v_o,v_2)$. Hence, $cos(v_o,v_1+v_2)<cos(v_o,v_1)$ and 
$cos(v_o,v_1+v_2)<cos(v_o,v_2)$. Therefore,
$$\frac{v_o(v_1+v_2)}{\mid v_1+v_2\mid}<\frac{v_ov_1}{\mid v_1\mid}\ \ \ \ \ 
and \ \ \ \ \ \frac{v_o(v_1+v_2)}{\mid v_1+v_2\mid}<\frac{v_ov_2}{\mid v_2\mid}$$
which conclude
\\
\\
$$2(\frac{v_o(v_1+v_2)}{\mid v_1+v_2\mid})<\frac{v_ov_1\mid v_2\mid+v_ov_2\mid v_1\mid}{\mid v_1\mid\mid v_2\mid}$$
$$2v_ov_1\mid v_1\mid\mid v_2\mid+2v_ov_2\mid v_1\mid\mid v_2\mid\ <\ v_ov_1\mid v_2\mid\mid v_1+v_2\mid+v_ov_2\mid v_1\mid\mid v_1+v_2\mid$$
$$v_ov_1\mid v_2\mid(2\mid v_1\mid-\mid v_1+v_2\mid)+v_ov_2\mid v_1\mid(2\mid v_2\mid-\mid v_1+v_2\mid)<0$$
However, 
$$0>v_ov_1\mid v_2\mid(2\mid v_1\mid-\mid v_1+v_2\mid)+v_ov_2\mid v_1\mid(2\mid v_2\mid-\mid v_1+v_2\mid)$$
$$\ge (min\{v_ov_1\mid v_2\mid,v_ov_2\mid v_1\mid\})(2)(\mid v_1\mid+\mid v_2\mid-\mid v_1+v_2\mid)\ge 0$$
which is a contradiction. Therefore, it is true for $n=3$. Suppose that it is true for $n=k-1$, and we want to prove it for $n=k$. For $t<k$, our inductive hypothesis implies that
$$\theta(v_o,w_1=\sum_{i=1}^{t}v_i)\le max\{\theta(v_o,v_i):\ i=1,...,t\}$$
$$\theta(v_o,w_2=\sum_{i=t+1}^kv_i)\le max\{\theta(v_o,v_i):\ i=t+1,...,k\}$$
Therefore,
$$\theta(v_o,\sum_{i=1}^kv_i)\le max\{\theta(v_o,w_i):\ i=1,2\}\le max\{\theta(v_o,v_i):\ i=1,...,k\}$$
and this completes the proof $\diamond$
\\
\\
\textbf{Theorem 8.} Based on the optimal solution of the SDP model (4), and by introducing $U=2v_o^*-v_a^*-v_b^*$, we have 
$$\mid U\mid =\sqrt{2},\ \ \ and\ \ \ \ \forall j\in V_\varepsilon :\ Uv_j^*=1$$
\\
\textbf{Proof.} 
$$\mid U\mid =\sqrt{UU}=\sqrt{4-1-1-1+1+0-1+0+1}=\sqrt{2}$$ 

Moreover, for each vertex $j$ in $V_\varepsilon$, we have 
$$v_o^*v_c^*+v_o^*v_j^*-v_c^*v_j^*=1\ \ \ c\in\{a,b\},\ \ j\in V_\varepsilon$$

Therefore, we obtain
$$v_c^*v_j^*=-0.5+v_o^*v_j^*\ \ \ c\in\{a,b\},\ \ j\in V_\varepsilon$$  
which concludes that $Uv_j^*=2v_o^*v_j^*-v_a^*v_j^*-v_b^*v_j^*=2v_o^*v_j^*+0.5-v_o^*v_j^*+0.5-v_o^*v_j^*=1$, and the angle between two vectors $U$ and $v_j^*$ is $45$ degrees for $j\in V_\varepsilon \ \diamond$  
\\

Now we can prove our main result.
\\
\\
\textbf{Theorem 9.} By solving the SDP model (4) on $G2$, it is impossible to have an optimal solution that meets the Property (1) 
on $G$, unless the induced graph on $V_\varepsilon$ is bipartite.
\\
\textbf{Proof.} Suppose that the optimal solution of the SDP model (4) meets the Property (1) on $G$. 
Therefore, for each edge $ij$ in $E_\varepsilon$ we have four normalized vectors $v_i^*,v_j^*,v_a^*,v_b^*$ which are almost perpendicular to each other, 
and a normalized vector $v_o^*$ with $0.5\leq v_o^*v_c\leq 0.5+\varepsilon$ $(c=i,j,a,b)$ which is almost equal to $0.5(v_i^*+v_j^*+v_a^*+v_b^*)$. In other words, $v_i^*+v_j^*$ is almost equal to $U=2v_o^*-v_a^*-v_b^*$ and there exists a vector $r_{i,j}$ with $v_i^*+v_j^*+r_{i,j}=U$, where $-\varepsilon\le \mid v_i^*+v_j^*\mid-\mid U\mid\le\varepsilon,\ \mid r_{i,j}\mid\leq \varepsilon$, and $cos(U,v_i^*+v_j^*)\ge 1-\varepsilon$.

If we have an odd cycle on $2t+1$ vertices in $G_\varepsilon =(V_\varepsilon,E_\varepsilon)$, then, by addition of the vectors in this cycle and introducing $W=(v_1+v_2)+(v_2+v_3)+...+$$
$$(v_{2t}+v_{2t+1})+(v_{2t+1}+v_1)$, we have 
$$U.W=\ 2(2t+1)=\mid U\mid \mid W\mid cos(U,W)$$
Where, $\theta(U,W)\le max\{\theta(U,v_1+v_2),...,\theta(U,v_{2t}+v_{2t+1}),\theta(U,v_{2t+1}+v_1)\}\le 1^o$.

By introducing $W'=\sum_{i=1}^{2t+1}v_i=\ 0.5W$,  the above summation can be written as follows,
$$U.W=2U.W'=2\mid U\mid \mid W'\mid cos(U,W')=\ 2(2t+1)$$
Where, $\theta(U,W')=\theta(U,W)\le 1^o$, and 
$$\frac{(t+0.5)\sqrt{2}}{1}\le\mid W'\mid =\frac{(t+0.5)\sqrt
{2}}{cos(U,W')}$$

By introducing $W_i=W'-v_i$, for $i=1,...,2t+1$, we have 
$$\mid v_i\mid^2=\mid W'\mid^2+\mid W_i\mid^2-2W'W_i\ \ \ \ \ (I)$$
and
$$\mid W'\mid\le\mid W_i\mid +1$$
and 
$$\sum_{i=1}^{2t+1}W_i=(2t+1)W'-\sum_{i=1}^{2t+1}v_i=2tW'$$
and
$$U.W_i=\ 2t=\mid U\mid \mid W_i\mid cos(U,W_i)$$
Where, $\theta(U,W_i)\le max\{\theta(U,v_{i+1}+v_{i+2}),...,\theta(U,v_{i-2}+v_{i-1})\}\le 1^o$, and
$$\frac{t\sqrt{2}}{1}\le\mid W_i\mid =\frac{t\sqrt{2}}{cos(U,W_i)}$$

By addition of the equation (I) in this cycle, we have 
$$\sum_{i=1}^{2t+1}\mid v_i\mid^2=\sum_{i=1}^{2t+1}\mid W'\mid^2+\sum_{i=1}^{2t+1}\mid W_i\mid^2-2\sum_{i=1}^{2t+1}W'W_i$$
Hence,
$$2t+1=(2t+1)\mid W'\mid^2+\sum_{i=1}^{2t+1}\mid W_i\mid^2-2W'\sum_{i=1}^{2t+1}W_i$$
$$=(2t+1)\mid W'\mid^2+\sum_{i=1}^{2t+1}\mid W_i\mid^2-2W'(2tW')$$
$$=(-2t+1)\mid W'\mid^2+\sum_{i=1}^{2t+1}\mid W_i\mid^2\ \ \ \ \ \ \ (II)$$

Let $\mid W_j\mid =max\{\mid W_i\mid\ :\ i=1,...,2t+1\}$. Then, we have 
$$(-2t+1)\mid W'\mid^2+\sum_{i=1}^{2t+1}\mid W_i\mid^2\le (-2t+1)(\mid W_j\mid +1)^2+(2t+1)\mid W_j\mid^2$$
$$\le 2\mid W_j\mid^2-4t\mid W_j\mid +2\mid W_j\mid -2t+1$$
$$=\frac{4t^2}{cos^2(U,W_j)}-\frac{4\sqrt{2}t^2}{cos(U,W_j)}+\frac{2\sqrt{2}t}{cos(U,W_j)}-2t+1$$
$$\le \frac{4t^2}{cos^2(1^o)}-\frac{4\sqrt{2}t^2}{cos(0^o)}+\frac{2\sqrt{2}t}{cos(1^o)}-2t+1$$
$$\le -1.6556t^2+0.8288t+1$$
However, $-1.6556t^2+0.8288t+1$ is less than $2t+1$, which contradicts the equation (II). 

Therefore, there is not any odd cycle in $G_\varepsilon$, and $G_\varepsilon$ is bipartite $\ \diamond$
\\

\textbf{proposition 2.} To produce a performance ratio of 1.999999 for problems with $z_{VCP}^*\geq\frac{n}{2}$, 
we should solve the SDP model (4) on $G2$. Then, the performance ratio of the solution is 1.999999, if it does not meet the Property (1). Otherwise, we can solve the VCP problem on the bipartite graph $G_\varepsilon$, 
where $\mid V_\varepsilon \mid\geq 0.989999n$, to produce a performance ratio of 1.999999. 
\\

Moreover, based on the Theorem (1) and the Proposition (2), to produce a performance ratio of 1.999999 for problems with $z_{VCP}^*<\frac{n}{2}$, it is sufficient to produce an extreme optimal solution for the LP relaxation of the model (1) and introducing $G2$ based on $G_{0.5}$. 
\\
\\
\textbf{Theorem 10.} The Optimal solution of the following LP model corresponds to an extreme optimal solution of the LP relaxation of the model (1).
$$(5)\ min_{s.t.}⁡z5=\sum_{i=1}^n (0.1)^i x_{i}$$
$$x_{i}+x_{j}\geq 1\ \ \ ij\in E$$
$$\sum_{i\in V} x_{i}=z^*_{LP\ relaxation\ of\ the\ model\ (1)}$$
$$0\leq x_{i}\leq +1\ \ \ i\in V$$
\\
\textbf{Proof.} The feasible region of the model (5) is an optimal face of the feasible region of the LP relaxation of the model (1), and its optimal solution corresponds to the solution of the following algorithm, based on the priority weights of the decision variables.
\\
\textbf{Step 0.} Let k=1 and $z^*$ be the optimal value of the LP relaxation of the model (1).
\\
\textbf{Step k.} Solve the following LP model.
\\
$$(6)\ min_{s.t.}z(k)=⁡x_{k}$$
$$x_{i}+x_{j}\geq 1\ \ \ ij\in E$$
$$\sum_{i\in V} x_{i}=z^*$$
$$x_i=x_i^*=z(k)^*\ \ \ i=1,\cdots , k-1$$
$$0\leq x_{i}\leq +1\ \ \ i\in V$$
\\
Let k=k+1. If $k<n$ repeat this step, otherwise, the solution $x^*$ is an extreme optimal solution of the LP relaxation of the model (1)$\ \diamond$ 
\\

Therefore, our algorithm to produce an approximation ratio of 1.999999, for arbitrary vertex cover problems, is as follows:
\\
\\
\textbf{Mahdis Algorithm} (To produce a vertex cover solution on graph G with a ratio factor $\rho \le 1.999999$)
\\
\textbf{Step 1.} Let $V^1=V^0=\{ \}$ and solve the LP relaxation of the model (1) on $G$.
\\
\textbf{Step 2.} If $z^*_{LP\ relaxation\ of\ the\ model\ (1)}<\frac{n}{2}$, then solve the LP model (5) 
to produce an extreme optimal solution of the LP relaxation of the model (1), and based on the solution  ($x_j^*\in \{0,0.5,1\}\ \ j\in V$), 
introduce $V^0=\{j\in V\mid x_j^*=0\}$, $V^{0.5}=\{j\in V\mid x_j^*=0.5\}$, $V^1=\{j\in V\mid x_j^*=1\}$, and let $G=G_{0.5}$ as the induced graph on the vertex set $V^{0.5}$.
\\
\textbf{Step 3.} Produce $G2$ based on $G$ and solve the SDP (4) model. 
\\
\textbf{Step 4.} If $\mid\{j\in V:\  v_o^*v_j^*<0.5 \} \mid > 0.000001n$ (the solution does not meet the Property (1.a)), then introduce $V_0=\{j\in V\mid v_o^*v_j^*<0.5\}$ and $V_1=V-V_0$ to produce a suitable solution $V_1\cup V_0$ which satisfies $\frac{\mid V_1 \mid}{z^*(G)}\leq 1.999999$. Then, go to Step 7. Otherwise, go to Step 5.
\\
\textbf{Step 5.} If $\mid\{j\in V:\  v_o^*v_j^*>0.5004 \} \mid > 0.01n$, then it is sufficient to produce an arbitrary 
VCP feasible solution $V=V_1\cup V_0$ to have $\frac{\mid V_1\mid}{z^*(G)}\leq 1.999999$ and go to Step 7. Otherwise, go to Step 6.
\\
\textbf{Step 6.} The solution meets the Property (1) and based on the Theorem (9), graph $G_\varepsilon$ is bipartite, 
and $\mid V_\varepsilon \mid\geq 0.989999n$. Therefore, solve the VCP problem on bipartite subgraph $G_\varepsilon$ and add all vertices of $V-V_\varepsilon$ to the solution to produce a feasible solution $V_1\cup V_0$ which satisfies $\frac{\mid V_1\mid}{z^*(G)}\leq 1.999999$. Then, go to Step 7.
\\
\textbf{Step 7.} The partitioning $(V_1\cup V^1)\cup (V_0\cup V^0)$ produces a VCP feasible solution on the original graph $G$ with an approximation ratio factor $\rho \le 1.999999$.
\\
\\
\textbf{proposition 3.} Based on the proposed 1.999999-approximation algorithm for the vertex cover problem, the unique games conjecture is not true.

\section{Conclusions}

One of the open problems regarding the vertex cover problem is the possibility of introducing an approximation algorithm within any constant factor better than 2. Here, we propose a new algorithm to produce a 1.999999-approximation ratio for the vertex cover problem on arbitrary graphs, which leads to the conclusion that the unique games conjecture is not true.
\\
\\
\textbf{Acknowledgment.} I would like to sincerely thank Dr. Flavia Bonomo for her useful comments and her endorsement for publishing in arXiv. 
\\
\\
\textbf{Competing Interest and Data Availability} 
\\
The authors have no relevant financial or non-financial interests to declare relevant to this article's content. Data sharing does not apply to this article as no data sets were generated or analyzed during the current study.
\\
\\
\textbf{References}
\\
1) Dinur I., Safra S. (2005) On the hardness of approximating minimum vertex cover. Annals of Mathematics, 162, 439-485.
\\
2) Khot S. (2002) On the power of unique 2-Prover 1-Round games. Proceeding of 34th ACM Symposium on Theory of Computing, STOC.
\\
3) Khot S., Regev O. (2008) Vertex cover might be hard to approximate to within $2-\varepsilon$. Journal of Computer and System Sciences, 74, 335-349.
\\
4) Nemhauser G.L., Trotter L.E. (1975) Vertex packing: Structural properties and algorithms. Mathematical Programming, 8, 232-248.

\end{document}